\documentstyle[12pt]{article}
\begin{document}
\title{ISSUES AND RAMIFICATIONS IN QUANTIZED FRACTAL SPACE TIME: AN INTERFACE
WITH QUANTUM SUPERSTRINGS}
\author{B.G. Sidharth$^*$\\
B.M. Birla Science Centre, Hyderabad 500 063 (India)}
\date{}
\maketitle
\footnotetext{$^*$E-mail:birlasc@hd1.vsnl.net.in}
\begin{abstract}
Recently a stochastic underpinning for space time has been considered, what
may be called Quantized Fractal Space Time. This leads us to a number of very
interesting consequences which are testable, and also provides a rationale
for several otherwise inexplicable features in Particle Physics and
Cosmology. These matters are investigated in the present paper.
\end{abstract}
\section{The Background ZPF}
We observe that the Bohm formulation discussed in detail in Chapter 3
converges to Nelson's stochastic formulation in
the context of the QMKNBH. Indeed Bohm's non local potential as also Nelson's
three conditions merely describe the QMKNBH as a vortex, the mass being given
by the self interaction, the radius of the vortex being the Compton wavelength.
\cite{r1}. We can get a clue to the origin of Quantum Mechanical fluctuations: Following Smolin we
observe that the non local stochastic theory becomes the classical local
theory in the thermodynamic limit, in which $N$ the number of particles
in the universe becomes infinitely large. However if $N$ is finite but
large, these fluctuations are of the order $1/\sqrt{N}$ of the dimensions
of the system, the universe in this case. Indeed as we will see in the nexr Section
this provides a holistic rationale for the "spooky"
non-locality of Quantum Theory.\\
We now remark that in the above formulation elementary particles, typically
electrons, can be thought of as 'twisted bits' of the electromagnetic
field.
Indeed it was pointed out by Barut and co-workers that wave packet solutions
of the mass less scalar fields appear as massive particles, while such
solutions for the electromagnetic field would provide a formulation of the
wave mechanics without assuming the Planck constant\cite{r2}. For example
this gives $E = l\omega$ rather than $E = \hbar \omega$, $l$ being
the angular momentum and $E$ the energy and $\omega$ the frequency.
Boudet\cite{r3}, also questions the necessity of the Planck
Constant. These theories
do not give a value to  Planck's constant
which merely appears as a proportionality factor, because all the equations
considered are linear. All this as also the zitterbewegung formulation of
Barut and Bracken and Hestenes described in Chapter 3, is superseded by the QMKNBH
theory and electrons
appear as twisted bits of the ZPF given by the relation, $E = \hbar \omega$,
instead of Barut's,
$E = l \omega$, where $l$ is the angular momentum. The question, whether this
characterizes the Planck constant, will be answered at the end of Section 5.
\section{Stochastic Conservation Laws}
Conservation Laws, as is universally known, play an important role in
Physics, starting with the simplest such laws relating to momentum and
energy. These laws provide rigid guidelines or constraints within which
physical processes take place.\\
These laws are observational, though a theoretical facade can be given by
relating them to underpinning symmetries\cite{r4}.\\
Quantum Theory, including Quantum Field Theory
is in conformity with the above picture. On the other hand the laws
of Thermodynamics have a different connotation: They are not rigid in the sense
that they are a statement about what is most likely to occur, or is an averaged
out statement.\\
In our formulation, the Compton wavelength represents a statistical uncertainty
(Cf. Chapter 6), given by
\begin{equation}
l \sim \frac{R}{\sqrt{N}}\label{e1}
\end{equation}
By now (\ref{e1}) is a familiar relation.
Given the above background we consider the following simplified EPR experiment,
discussed elsewhere\cite{r5}.\\
Two structureless and spinless particles which are initially together, for
example in a bound state get separated and move in opposite directions along
the same straight line. A measurement of the momentum of one of the particles,
say $A$ gives us immediately the momentum of the other particle $B$. The
latter is equal and opposite to the former owing to the conservation law
of linear momentum. It is surprising that this statement should be true
in Quantum Theory also because the momentum of particle $B$ does not
have an apriori value, but can only be determined by a separate acausal experiment
performed on it.\\
This is the well known non locality inherent in Quantum Theory. It ceases to be mysterious
if we recognize the fact that the conservation of momentum is itself a non
local statement because it is a direct consequence of the homogeneity of
space as we will see again in the next Chapter: Infact the displacement operator $\frac{d}{dx}$ is, given the
homogeneity of space, independent of $x$ and this leads to the conservation
of momentum in Quantum Theory (cf.ref.\cite{r6}). The displacement $\delta x$ which gives rise
to the above displacement operator is an instantaneous shift of origin
corresponding to an infinite velocity and is compatible with a closed system.
It is valid if the instantaneous displacement can also be considered to be
an actual displacement in real time $\delta t$. This happens for stationary
states, when the overall energy remains constant.\\
It must be borne in mind that the space and time displacement operators are
on the same footing only in this case\cite{r7}. Indeed in relativistic Quantum
Mechanics, $x$ and $t$ are put on the same footing -  but special relativity
itself deals with inertial, that is relatively unaccelerated frames.\\
Any field theory deals with different points at the same instant
of time. But if we are to have information about different points, then given
the finite velocity of light, we will get this information at different
times. All this information can refer to the same instant of time only in
a stationary situation. We will return to this point. Further the field equations are obtained by a
suitable variational principle,
\begin{equation}
\delta I = 0\label{e2}
\end{equation}
In deducing these equations, the $\delta$ operator which corresponds to an
arbitrary variation, commutes with the space and time derivatives, that is
the momentum and energy operators which in our picture constitute a complete
set of observables. As such the apparently arbitrary operator $\delta$ in
(\ref{e2}) is constrained to be a function of these (stationary) variables\cite{r8}.\\
All this underscores two facts: First we implicitly consider an apriori homogenous space,
that is physical space. Secondly though we consider in the relativisitic
picture the space and time coordinates to be on the same footing, infact
they are not as pointed out by Wheeler\cite{r9}. Our understanding or perception of the universe is based on
"all space (or as much of it as possible) at one instant of time".\\
However, in conventional theory this is at best an approximation. Moreover
in our formulation, the particles are fluctuationally created out of
a background ZPF, and, it is these $N$ particles that define physical space,
which is no longer apriori as in the Newtonian formulation. It is only in the thermodynamic limit in which
$N \to \infty$ and $l \to 0$, in (\ref{e1}), that we recover the above
classical picture of a rigid homogenous space, with the conservation laws.\\
In other words the above conservation laws are strictly valid in the
thermodynamic limit, but are otherwise approximate, though very nearly correct
because $N$ is so large.\\
Our formulation leads to a cosmology in which $\sqrt{N}$ particles
are fluctuationally created from the background ZPF (Cf. Chapter 7),
so that the violation of energy conservation is proportional
to $\frac{1}{\sqrt{N}}$. From (\ref{e1}) also we could similarly infer that
the violation of momentum conservation is proportional to $\frac{1}{\sqrt{N}}$
(per particle).\\
All this implies that there is a small but non-zero
probability that the measurement of the particle $A$ in the above experiment
will not give information about the particle $B$.\\
This last conclusion has also been drawn by Gaeta\cite{r10} who considers a background
Brownian or Nelson-Garbaczewski-Vigier noise(the ZPF referred to above) as sustaining Nelson's Stochastic
Mechanics (and the Schrodinger equation).\\
In conclusion, the conservation laws of Physics are conservation laws in the
thermodynamic sense.
\section{Quantized Space Time, Time's Arrow and Parity Breakdown}
The arrow of time has been a puzzle for a long time. As is well known, the
laws of Newtonian Mechanics, Electromagnetism or Quantum Theory do not
provide an arrow of time - they are equally valid under time reversal, with
only one exception. This is in the well known problem of Kaon decay. On the
other hand it is in Thermodynamics and Cosmology that we find an arrow of
time \cite{r11}. Indeed it has been shown that stochastic processes are needed for
irreversibility\cite{r12}.\\
It is also true that there has been no theoretical rationale for the Kaon
puzzle which we will touch upon shortly. We will try to find such a
theoretical understanding in the context of our quantized space-time, $\sim \hbar /
(\mbox{energy})$, that is the Compton time\cite{r13}.\\
Let us start with one of the simplest quantum mechanical systems, one which can
be in either of two sates separated by a small energy\cite{r14}.
The system flips from one state to another unpredictably and this "life time"
and the energy spread satisfy the Uncertainity Principle, so that the former
is a Compton time. We have:
\begin{equation}
\imath \hbar \frac{d\psi_\imath}{dt} \approx \imath \hbar [\frac{\psi_\imath
(t+n\tau)-\psi_\imath (t)}{n\tau}] = \sum^{2}_{\imath =1} H_{\imath j}
\psi_\imath\label{e3}
\end{equation}
$$\psi_\imath  = e^{\frac{\imath}{\hbar}Et} \phi_\imath$$
where, $H_{11} = H_{22}$ (which we set $= 0$ as only relative energies of the
two levels are being considered) and $H_{12} = H_{21} =E$,
by symmetry. Unlike in the usual theory where $\delta t = n \tau \to 0$, in the
case of quantized space-time $n$ is a positive integer. So the second
term of (\ref{e3}) reduces to
\begin{equation}
[E + \imath \frac{E^2 \tau}{\hbar}] \psi_\imath = [E (1+\imath )]\psi_\imath,
\mbox{as} \quad \tau = \hbar/E\label{e4}
\end{equation}
Interestingly, in the above analysis, in (\ref{e4}), the fact
that the real and imaginary parts are of the same order is infact
borne out by experiment.\\
From (\ref{e3}) we see that the Hamiltonian is not Hermitian that is it
admits complex Eigen values indicative of decay, if the life times of
the states are $\sim \tau$.\\
In general this would imply the exotic fact that if a state starts out as
$\psi_1$ and decays, then there would be a non zero probability of seeing
in addition the decay products of the state $\psi_2$. In the process it is
possible that some symmetries which are preserved in the decay of $\psi_1$
or $\psi_2$ separately, are voilated.\\
In this context we will now consider the Kaon puzzle. As is well known from
the original work of Gellmann and Pais, the two state analysis above is
applicable here\cite{r15,r16}. In the words of Penrose\cite{r17}, "the tiny fact of an
almost completely hidden time-asymmetry seems genuinely to be present in
the $K^0$-decay. It is hard to believe that nature is not, so to speak,
trying to tell something through the results of this delicate and beautiful
experiment." On the other hand as Feynman put it\cite{r18}, "if there is
any place where we have a chance to test the main principles of quantum
mechanics in the purest way.....this is it."\\
What happens in this well
known problem is, that given $CP$ invariance, a beam of $K^0$ masons can
be considered to be in a two state system as above, one being the short lived component
$K^S$ which decays into two pions and the other being the long lived state
$K^L$ which decays into three pions. In this case $E \sim 10^{10} \hbar$\cite{r15},
so that $\tau \sim 10^{-10}sec$. After a lapse of time greater than
the typical decay period, no two pion decays should be seen in a beam consisting
initially of the $K^0$ particle. Otherwise there would be violation of
$CP$ invariance and therefore also $T$ invariance. However exactly this
violation was observed as early as 1964\cite{r19}. This violation of time
reversal has now been confirmed directly by experiments at
Fermilab and CERN\cite{r20}.\\
We would like to point out that the Kaon puzzle has a natural explanation
in the quantized time scenario discussed above.
Further, we have shown that
the discreteness leads to the non commutative geometry
\begin{equation}
[x,y] = 0 (l^2), [x,p_x]=\imath \hbar [1+l^2]\label{e5}
\end{equation}
and similar equations. If terms $\sim l^2$ are neglected we get back the usual
Quantum Theory. However retaining these terms, we deduced in Chapter 6 the
Dirac equation. Moreover it can be seen that given (\ref{e5}) space
rerlection symmetry no longer holds. This violation is an $O(l^2)$ effect.\\
This is not surprising. It has already been pointed out that the
space time divide viz., $x + \imath ct$ arises due to the zitterbewegung
or double Weiner process in the Compton wavelength - and in this derivation
terms $\sim (ct)^2 \sim l^2$ were neglected. However if these terms are
retained, then we get a correction to the usual theory including special
relativity. (We will come back to this point shortly.)\\
To see this more clearly let us as in Chapter 3
as a first approximation treat the continuum as a series of discrete points
separated by a distance $l$, which then leads to
\begin{equation}
Ea(x_n) = E_0a(x_n) - Aa(x_n + l) - Aa(x_n - l)\label{e6}
\end{equation}
When $l$ is made to tend to zero, it was shown that from (\ref{e6}) we recover the
Schrodinger equation, and further, we have,
\begin{equation}
E = E_0 - 2A cos kl.\label{e7}
\end{equation}
The zero of energy was chosen such that $E = 2A = mc^2$, the rest energy of the
particle, in the limit $l \to 0$. However if we retain terms
$\sim l^2$, then from (\ref{e7}) we will have instead
\begin{equation}
\left| \frac{E}{mc^2} - 1\right| \sim 0 (l^2)\label{e8}
\end{equation}
Equation (\ref{e8}) shows the correction to the energy mass formula, where again we
recover the usual formula in the limit $O(l^2) \approx 0$.\\
It must be mentioned that all this would be true in principle for discrete
space time, even if the minimum cut off was not at the Compton scale.\\
Intuitively this should be obvious: Space time reflection symmetries are
based on a space time continuum picture.\\
Let us now consider some further imprints of discrete space time\cite{r21}.\\
First we consider the case of the neutral pion. As we saw in Chapter 4, this pion decays
into an electron and a positron. Could we think
of it as an electron-positron bound state also?
In this case we have,
\begin{equation}
\frac{mv^2}{r} = \frac{e^2}{r^2}\label{e9}
\end{equation}
Consistently with the above formulation, if we take $v = c$ from (\ref{e9}) we get the correct Compton wavelength
$l_\pi = r$ of the pion.\\
However this appears to go against the fact that there would be pair annihilation
with the release of two photons. Nevertheless if we consider discrete space time,
the situation would be different. In this case the Schrodinger equation
\begin{equation}
H \psi = E \psi\label{e10}
\end{equation}
where $H$ contains the above Coulumb interaction could be written, in terms
of the space and time separated wave function components as,
\begin{equation}
H\psi = E \phi T = \phi \imath \hbar [\frac{T(t-\tau)-T}{\tau}]\label{e11}
\end{equation}
where $\tau$ is the minimum time cut off which in the above work has been taken to be the Compton
time. If, as usual we let $T = exp (irt)$ we get
\begin{equation}
E = -\frac{2\hbar}{\tau} sin \frac{\tau r}{2}\label{e12}
\end{equation}
(\ref{e12}) shows that if,
\begin{equation}
| E | < \frac{2\hbar}{\tau}\label{e13}
\end{equation}
holds then there are stable bound states. Indeed inequality (\ref{e13}) holds
good when $\tau$ is the Compton time and $E$ is the total energy $mc^2$. Even if
inequality (\ref{e13}) is reversed, there are decaying states which are relatively
stable around the cut off energy $\frac{2\hbar}{\tau}$.\\
This is the explanation for treating the pion as a bound state of an electron
and a positron, as indeed is borne out by its decay mode.
The situation is similar to
the case of Bohr orbits-- there also the electrons would according to classical
ideas have collapsed into the nucleus and the atoms would have disappeared. In
this case it is the discrete nature of space time which enables the pion to be
a bound state as described by (\ref{e9}).
\section{Magnetic Effects}
If as discussed in Chapter 3 and subsequently, the electron is indeed a Kerr-Newman type charged black hole, it can be
approximated by a solenoid and we could expect an Aharonov-Bohm type of
effect, due to the vector potential $\vec A$ which would give rise to shift
in the phase in a two slit experiment for example\cite{r22}. This shift
is given by
\begin{equation}
\Delta \delta_{\hat B} = \frac{e}{\hbar} \oint \vec A . \vec{ds}\label{e14}
\end{equation}
while the shift due to the electric charge would be
\begin{equation}
\Delta \delta_{\hat E} = -\frac{e}{\hbar} \int A_0 dt\label{e15}
\end{equation}
where $A_0$ is the electrostatic potential. In the above formulation
we would have
\begin{equation}
\vec A \sim \frac{1}{c} A_0\label{e16}
\end{equation}
Substitution of (\ref{e16}) in (\ref{e14}) and (\ref{e15}) shows that
the magnetic effect $\sim \frac{v}{c}$ times the electric effect.\\
Further, the magnetic component of a Kerr-Newman black hole, as we saw in
Chapter 3 is given by
\begin{equation}
B_{\hat r} = \frac{2ea}{r^3} cos \Theta + 0(\frac{1}{r^4}), B_{\hat \Theta} =
\frac{ea sin\Theta}{r^3} + 0(\frac{1}{r^4}), B_{\hat \phi} = 0,\label{e17}
\end{equation}
while the electrical part is
\begin{equation}
E_{\hat r} = \frac{e}{r^2} + 0(\frac{1}{r^3}), E_{\hat \Theta} = 0(\frac{1}{r^4}),
E_{\hat \phi} = 0,\label{e18}
\end{equation}
Equations (\ref{e17}) and (\ref{e18}) show that in addition to the usual
dipole magnetic field, there is a shorter range magnetic field given by terms
$\sim \frac{1}{r^4}$. In this context it is interesting to note that an extra so called
$B^{(3)}$ magnetic field of shorter range and probably mediated by massive photons
has indeed been observed and studied over the past few years\cite{r23}.
\section{Stochastic Holism and the Number of Arbitrary Parameters}
The discrete space time or zitterbewegung has an underpinning that is
stochastic. The picture leads to the goal of Wheeler's 'Law without Law'
as we saw in Chapter 6.
Furthermore the picture that emerges is Machian. This is evident
from equations like (\ref{e2}), (\ref{e3}) and (\ref{e6}) of Chapter 7-- the micro depends
on the macro. So the final picture that emerges is one of stochastic holism.\\
Another way of expressing the above point is by observing that the
interactions are relational. For example, in the equation leading to
(\ref{e7}), of Chapter 7,
if the number of particles in the universe tends to $1$, then as we saw in Chapter 4, the
gravitational and electromagnetic interactions would be equal, this happening
at the Planck scale, where the Compton wavelength equals the Schwarzchild
radius\cite{r24}.\\
Infact as was shown in Chapter 6, when $N$ the number of
particles in the universe is $1$ we have a Planck particle with a
short life time $\sim 10^{-42}secs$ due to the Hawking radiation but with
$N \sim 10^{80}$ particles as in the present universe we have the pion as
the typical particle with a stable life time $\sim$ of the age of the
universe due to the Hagedorn on radiation.\\
Let us now consider the following aspect\cite{r25}:
It is well known that there are 18 arbitrary parameters in contemporary
physics. We on the other hand have been working with the micro physical
constants referred to earlier viz., the electron (or pion) mass or Compton
wavelength, the Planck constant, the fundamental unit of charge and the velocity
of light. These along with the number of particles $N$ as the only free parameter
can generate the mass, radius and age of the universe as also the Hubble
constant.\\
If we closely look at the equations (\ref{e11}) of Chapter 4 or (\ref{e7})
of Chapter 7, giving the gravitational and electromagnetic strength ratios, we can actually deduce the relation,
\begin{equation}
l = \frac{e^2}{mc^2}\label{e19}
\end{equation}
In other words we have deduced the pion mass in terms of the electron mass, or,
given the pion mass and the electron mass, we have deduced the fine structure
constant. From the point of view of the order of magnitude theory in which
the distinction between the electron, pion and proton gets blurred, what
equation (\ref{e19}) means is, that the Planck constant itself depends on
$e$ and $c$ (and $m$). Further in the Kerr-Newman type characterisation of
the electron, in Chapter 3 the charge $e$ is really equivalent to the spinorial tensor
density $(n=1)$. In this sense $e$ also is pre determined
and we are left with a minimum length viz. the Compton length and a minumum time
viz. the Compton time (or a maximal velocity $c$) as the only fundamental
microphysical constants.\\
Let us try to further refine this line of thought.
We observe that a discrete space time picture leads to the non commutative
geometry alluded to earlier (\ref{e5}).\\
Infact we would have in this case, more fully,
\begin{equation}
[x,y] = 0(l^2),[x,p_x] = \imath \hbar [1 + l^2], [t,E] = \imath \hbar [1+\tau^2]\label{e20}
\end{equation}
What (\ref{e20}) means is that there is a higher order correction to the
Heisenberg Uncertainity Principle. Infact from (\ref{e20}) we can easily
conclude that there is an extra energy $E'$ given by
\begin{equation}
\frac{E'}{mc^2}\sim \tau^2 \sim l^4 \sim \frac{1}{\sqrt{N}}\label{e21}
\end{equation}
In (\ref{e21}), the appearance of $\frac{1}{\sqrt{N}}$ where $N$ is the number
of particles in the universe appears at first sight to be purely accidental:
We have not deduced it. However this is not so. Infact from the picture of the fluctuational
creation of particles alluded to in section 2, we get
\begin{equation}
\frac{E'}{mc^2} \sim \frac{1}{\sqrt{N}}\label{e22}
\end{equation}
It can be seen that (\ref{e22}) and (\ref{e21}), deduced from two totally
different standpoints, are infact the same.
A consequence is the following fact: We have just seen
that the micro physical constants namely an elementary particle mass, for example
the electron mass $m$ (or Compton wavelength), a universal maximal velocity $c$ together with $N$
the number of particles in the universe were the only free parameters or
arbitrary constants. From (\ref{e21}) we can see that there is a further
narrowing down to just two arbitrary parameters, for example the maximal
velocity $c$ and $N$. Given these two, the microphysical constants, including
the Planck constant can be characteriized, thus answering the question at the
end of section 1. It must be emphasized that what is required is a universal
maximum velocity in principle - its exact value is not important. Then, $N$
becomes the only parameter! All this is very much in the spirit of Feynman's
quotation in Chapter 1 as also the ancient Upanishadic tradition of seeing
nature as different aspects of one phenomenon.
\section{The Origin of a Metric}
We first make a few preliminary remarks. When we talk of a metric or the distance
between two "points" or "particles", a concept that is implicit is that of
topological "nearness" - we require an underpinning of a suitably large number
of "open" sets\cite{r26}. Let us now abandon the absolute or background space
time and consider, for simplicity, a universe (or set) that consists solely
of two particles. The question of the distance between these particles (quite
apart from the question of the observer) becomes meaningless. Indeed, this is
so for a universe consisting of a finite number of particles. For, we
could isolate any two of them, and the distance between them would have no
meaning. We can intuitively appreciate that we would infact need distances of intermediate points. So for a
meaningful distance, the concepts of open sets, connectedness and the like
reenter in which case such an isolation would not be possible.\\
More formally let us define a neighbourhood of a particle (or point) A of a set of
particles as a subset which contains A and atleast one other distinct
element. Now, given two particles (or points) A and B, let us consider a
neighbourhood containing both of them, $n(A,B)$ say. We require a non-empty
set containing atleast one of A and B and atleast one other particle C, such
that $n(A,B)\supset n(A,C)$, and so on. Strictly, this "nested" sequence
should not terminate. For, if it does, then we end up with a set $n(A,P)$ consisting
of two isolated "particles" or points, and the "distance" $d(A,P)$ is meaningless.
For practical purposes, in the spirit of Wheeler's approximation, this sequence
has to be very large.\\
Such an approximation has an immediate application. Our universe consists of some
$N \sim 10^{80}$ particles (or points), each point being "defined" within the
Compton wavelength $l$. Inside $l$, space time in the usual sense breaks down -
we have the unphysical zitterbewegung effects. Indeed $l$ for a Planck
particle of mass $\sim 10^{-5}gm$ is precisely the Planck scale.\\
We now assume the following property\cite{r27}: Given two distinct elements (or even
subsets) $A$ and $B$, there is a neighbourhood $N_{A_1}$ such that $A$
belongs to $N_{A_1}, B$ does not belong to $N_{A_1}$ and also given any
$N_{A_1}$, there exists a neighbourhood $N_{A_\frac{1}{2}}$ such that
$A \subset N_{A_\frac{1}{2}} \subset N_{A_1}$, that is there exists an infinite sequence of
neighbourhoods between $A$ and $B$. In other words we introduce topological
closeness.\\
From here, as in the derivation of Urysohn's lemma\cite{r26}, we could define
a mapping $f$ such that $f (A) = 0$ and $f(B) = 1$ and which takes on all
intermediate values. We could now define a metric, $d(A,B) = |f(A)-f(B)|.$
We could easily verify that this satisfies the properties of a metric.\\
It must be remarked that the metric turns out to be again, a result of a global
or a series of larger sets, unlike the usual local picture in which it is the
other way round.
\section{Kaluza-Klein Theories and Quantized Super Strings}
In Chapter 1, we briefly alluded to string theory. Though our subsequent
considerations were in a different class, there is a surprising interface,
as we will now see.
Our starting point is the fact encountered in Chapter 6 that the fractal dimension of a Brownian quantum
path is 2.
This was further analysed  and it was explained
that this is symptomatic of Quantized Fractal space time and it was shown that
infact the coordinate $x$ becomes $x+\imath ct$. The
complex coordinates or equivalently non-Hermitian position operators are
symptomatic of the unphysical zitterbewegung which is eliminated after an averaging
over the Compton scale. In this picture the fluctuational creation of particles was
taken into account in a consistent cosmological scheme in Chapter 7.\\
It is well known that the generalization of the complex $x$ coordinate to three
dimensions leads to quarternions\cite{r28}, and the Pauli spin matrices.\\
We next return to the model of an electron as a Quantum Mechanical Kerr-Newman Black Hole.
Infact in Chapter 3, we deduced electromagnetism in two
ways. The first was by considering an imaginary shift,
\begin{equation}
x^\mu \to x^\mu + \imath a^\mu , (a^\mu \sim \mbox{Compton scale})\label{e23}
\end{equation}
in a Quantum Mechanical context. This lead to
\begin{equation}
\imath \hbar \frac{\partial}{\partial x^\mu} \to \imath \hbar \frac{\partial}
{\partial x^\mu} + \frac{\hbar}{a^\mu}\label{e24}
\end{equation}
and the second term on the right side of (\ref{e24}) was shown to be the
electromagnetic vector potential $A^\mu$,
\begin{equation}
A^\mu = \hbar/a^\mu\label{e25}
\end{equation}
The second was by taking into account the fact that at the Compton scale, it
is the so called negative energy two spinors $\chi$ of the Dirac bispinor that
dominate where,
$$\chi \to -\chi$$
under reflections. This lead to the tensor density property,
\begin{equation}
\frac{\partial}{\partial x^\mu} to \frac{\partial}{\partial x^\mu} - \Gamma_\nu^{\mu \nu}\label{e26}
\end{equation}
the second term on the right side of (\ref{e26}) being identified with $A^\mu$,
\begin{equation}
A^\mu = \hbar \Gamma_\nu^{\mu \nu}\label{e27}
\end{equation}
It was pointed out that (\ref{e27}) is formally and mathematically identical to
Weyl's original formulation, except that here it arises due to the
purely Quantum Mechanical spinorial behaviour whereas Weyl had put it by hand.\\
Another early scheme for the unification of gravitation and electromagnetism
as referred to earlier was that put forward by Kaluza and Klein\cite{r29,r30,r31} in which
an extra dimension was introduced and taken to be curled up. This idea has
resurfaced in recent years in String Theory.\\
We will first show that the characterization of $A^\mu$ in (\ref{e25}) is
identical to a Kaluza Klein formulation. Then we will show that equations (\ref{e26})
and (\ref{e27}) really denote the fact that the geometry around an electron is
non-integrable. Finally we will show that infact both (\ref{e24}) or (\ref{e25})
and (\ref{e26}) or (\ref{e27}) are the same formulations (as can be guessed
heuristically by comparing (\ref{e24}) and (\ref{e26})).\\
We first observe that the transformation (\ref{e23}) can be written as,
\begin{equation}
x^\imath \to x^\imath + \alpha^{\imath 5} x_5\label{e28}
\end{equation}
where $\alpha_{\imath 5}$ in (\ref{e28}) will represent a small shift from the
Minkowski metric $g_{\imath j}$, and $\imath , j=1,2,3,4,5,x^5$ being a fifth
coordinate introduced for purely mathematical conversion.\\
Owing to (\ref{e28}), we will have,
\begin{equation}
g_{\imath j} dx^\imath dx^j \to g_{\imath j} dx^\imath dx^j + (g_{\imath j}
\alpha^{j5}) dx^\imath dx_5\label{e29}
\end{equation}
In Kaluza's formulation,
\begin{equation}
A_\mu \propto g_{\mu 5}\label{e30}
\end{equation}
Comparison of (\ref{e30}), (\ref{e28}) and (\ref{e29}) with (\ref{e23}) and (\ref{e25}) shows
that indeed this is the case. That is, the formulation given in (\ref{e23})
and (\ref{e24}) could be thought of as introducing a fifth curled up
dimension, as in the Kaluza-Klein theory.\\
To see why the Quantum Mechanical formulation (\ref{e26}) and (\ref{e27})
corresponds to Weyl's theory, we start with the effect of an infinitesimal
parallel displacement of a vector\cite{r32}.
\begin{equation}
\delta a^\sigma = -\Gamma^\sigma_{\mu \nu} a^\mu d x^{\nu}\label{e31}
\end{equation}
As is well known, (\ref{e31}) represents the extra effect in displacements,
due to the curvature of space - in a flat space, the right side would vanish.
Considering partial derivatives with respect to the $\mu^{th}$
coordinate, this would mean that, due to (\ref{e31})
\begin{equation}
\frac{\partial a^\sigma}{\partial x^\mu} \to \frac{\partial a^\sigma}{\partial x^\mu}
- \Gamma^\sigma_{\mu \nu} a^\nu\label{e32}
\end{equation}
The second term on the right side of (\ref{e32}) can be written as:
$$-\Gamma^\lambda_{\mu \nu} g^\nu_\lambda a^\sigma = -\Gamma^\nu_{\mu \nu} a^\sigma$$
where we have utilized the property that in the above formulation as seen in Chapter 3,
$$g_{\mu \nu} = \eta_{\mu \nu} + h_{\mu \nu},$$
$\eta_{\mu \nu}$ being the Minkowski metric and $h_{\mu \nu}$ a small
correction whose square is neglected.\\
That is, (\ref{e32}) becomes,
\begin{equation}
\frac{\partial}{\partial x^\mu} \to \frac{\partial}{\partial x^\mu} -
\Gamma^\nu_{\mu \nu}\label{e33}
\end{equation}
The relation (\ref{e33}) is the same as the relation (\ref{e26}).\\
We will next show the correspondence between (\ref{e33}) or (\ref{e27})
or (\ref{e26}) and (\ref{e25}) or (\ref{e24}). To see this simply we note that the
geodesic equation is,
\begin{equation}
\dot u^\mu \equiv \frac{du^\mu}{ds} = \Gamma^\mu_{\nu \sigma} u^\nu u^\sigma\label{e34}
\end{equation}
We also use the fact that in the Quantum Mechanical Kerr-Newman Black Hole model referred
to, we have as in Chapter 3
$$u^\mu = c \quad \mbox{for}\quad \mu = 1, 2 \quad \mbox{and}3,$$
while,
$$|\dot u^\mu | = |u^\mu | \frac{mc^2}{\hbar}$$
So, from (\ref{e34}) we get,
$$\Gamma^\mu_{\nu \mu} = \frac{1}{a^\nu} , |a^\nu | = \frac{\hbar}{mc}$$
This establishes the required identity.\\
We now come to the interface with Quantum Super Strings.
We have already seen that the Quantized Fractal space time referred to really
leads to a non-commutative geometry, given by (\ref{e20})
It was also seen that these relations directly lead to the Dirac equation:
Quantized Fractal space time is
the underpinning for Quantum Mechanical spin or the Quantum Mechanical Kerr-Newman Black
Hole, that is ultimately equations like (\ref{e24}) or (\ref{e25}) and
(\ref{e26}) or (\ref{e27}).\\
It is also true that both the Kaluza Klein formulation and the non commutative
geometry (\ref{e20}) hold in the theory of Quantum Superstrings (QSS).\\
Infact we get from here a clue to the mysterious six extra curled up dimensions
of Quantum Superstring Theory. For this we observe that (\ref{e20}) gives
an additional contribution to the Heisenberg Uncertainity Principle and we
can easily deduce
$$\Delta p \Delta x \sim \hbar l^2$$
Remembering that at this Compton scale
$$\Delta p \sim mc$$
It follows that
\begin{equation}
\Delta x \sim l^3\label{e35}
\end{equation}
as $l \sim 10^{-11}cms$ for the electron we recover from (\ref{e35}) the
Planck Scale, as well as a rationale for the peculiar fact that the Planck
Scale is the cube of the electron Compton scale.\\
More importantly, what (\ref{e35}) shows is, that at this level, the single
dimension along the $x$ axis shows up as being three dimensional. That is there
are two extra dimensions, in the unphysical region below the Compton scale.
As this is true for the $y$ and $z$ coordinates also, there are a total of six
curled up or unphysical or inaccessible dimensions in the context of the
preceding section.\\
If we start with equations (\ref{e23}) to (\ref{e25}) which are related to QFST
(Quantized Fractal space time) and the non-commutative relation (\ref{e20})
we obtain a unification of electromagnetism and gravitation. On the other hand
if we consider the spinorial behaviour of the Dirac wave function, we get
(\ref{e26}) or (\ref{e27}). The former has been seen to be the same as the
Kaluza formulation while the latter is formally similar to the Weyl formulation -
but in this case (\ref{e27}) is not put in by hand. Rather it is a Quantum
Mechanical consequence. We have thus shown that these two approaches are the
same. The extra dimensions are thus seen to be confined to the unphysical
Compton scale - classically speaking they are curled up or inaccessible.\\
In a sense this is not surprising. The bridge between the two approaches was
the Kerr-Newman metric which uses, though without a clear physical meaning in
classical theory, the transformation (\ref{e23}). The reason why an imaginary
shift is associated with spin is to be found in the Quantum Mechanical
zitterbewegung and the consequent QFST.\\
Wheeler remarked as quoted in Chapter 4 \cite{r9}, "the most evident shortcoming of the geometrodynamic
model as it stands is this, that it fails to supply any completely natural
place for spin $1/2$ in general and for the neutrino, in particular", while
"it is impossible to accept any description of elementary particles that does
not have a place for spin half." Infact the bridge between the two is the
transformation (\ref{e23}). It introduces spin half into general relativity
and curvature to the electron theory, via the equation (\ref{e27}) or (\ref{e32}).\\
In this context it is interesting to note that El Naschie has given the fractal
formulation of gravitation\cite{r33}.\\
Thus apparently disparate concepts like the Kaluza Klein and Weyl formulations, Quantum Mechanical Black
Holes, Quantized Fractal space time and QSS are seen to have a harmonius
overlap, in the context of QFST with its roots in the fluctuational creation
of particles\cite{r34,r35}.\\
It is worth pointing out some of the similarities between String theories and
our formulation. The former started off, by considering one dimensional
extended objects or strings, the extension being of the order of the proton
Compton wavelength, vibrating and rotating with the speed of light (Cf. refs.
given in Chapter 1). Not only could space time points and singularities
be fudged, but further the angular momenta were proportional to the squares
of the masses, defining the well known Regge trajectories, as also in our
formulation (Cf. Chapter 12, Equation (\ref{e14})). All this is not
surprising.\\
In particular, QSS deals with Planck length phenomena, the Kaluza-Klein curled
up extra dimensions and leads to the non commutative geometry (\ref{e20}). QFST
on the other hand, deals with phenomena at the Compton scale, space time being
unphysical below this scale. Yet it leads us back to the Planck scale, the
same number of extra, curled up, Kaluza-Klein dimensions and the same non-
commutative geometry (\ref{e20}), once the meaning of (\ref{e35}) (or the
modification of the Heisenberg Uncertainity Principle) is recognized. In this
interpretation, the situation is similar to the fractal one dimensioinal
Brownian path becoming two (or three) dimensional. The key is the transformation
(\ref{e23}), which we first encountered right at the beginning, in Chapter
3 itself. It conceals zitterbewegung, leads to the Kerr-Newman metric, QFST
and what not!\\
Finally it is worth emphasizing that both in Strings and in our formulation
the Compton wavelength extension provides a rationale for the dual resonance
model, which originated from the Regge trajectories and then gave the initial
motivation for String theory.
\section{Resolution, Unification and the Core of the Electron}
El Naschie\cite{r36} has referred to the fact that there is no apriori fixed length scale (the
Biedenharn conjecture). Indeed it has been argued in the above
context that depending on our scale of resolution, we encounter
electromagnetism well outside the Compton wavelength, strong interactions at the
Compton wavelength or slightly below it and only gravitation at the Planck
scale. The differences between the various interactions are a manifestation
of the resolution.\\
In this connection it may be noted that we can refer to the core of the electron
$\sim 10^{-20}cms$, as indeed has been experimentally noticed by Dehmelt and
co-workers\cite{r37}. It is interesting that this can be deduced in the
context of the electron as a Quantum Mechanical Kerr-Newman Black Hole.\\
It was shown in Chapter 3 that for distances of the order of the Compton
wavelength the potential is given in its QCD form
\begin{equation}
V \approx - \frac{\beta M}{r} + 8\beta M (\frac{Mc^2}{\hbar})^2 .r\label{e36}
\end{equation}
For small values of $r$ the potential (\ref{e36}) can be written as
\begin{equation}
V \approx \frac{A}{r} e^{-\mu^2 r^2}, \quad \mu = \frac{Mc^2}{\hbar}\label{e37}
\end{equation}
It follows from (\ref{e37}) that
\begin{equation}
r \sim \frac{1}{\mu} \sim 10^{-21}cm.\label{e38}
\end{equation}
Curiously enough in (\ref{e37}), $r$ appears as a time, which is to be expected
because at the horizon of a black hole $r$ and $t$ interchange roles.\\
One could reach the same conclusion, as given in equation (\ref{e38}) from
a different angle. In the Schrodinger equation which is used in QCD, with the potential given
by (\ref{e36}), one could verify that the wave function is of the type
$f(r).e^{-\frac{\mu r}{2}}$, where the same $\mu$ appears in (\ref{e37}). Thus,
once again we have a wave packet which is negligible outside the distance
given by (\ref{e38}).\\
It may be noted that Brodsky and Drell\cite{r38} had suggested from a very
different viewpoint viz., the anomalous magnetic moment of the electron,  that
its size would be limited by
$10^{-20}cm$. The result (\ref{e38}) as pointed out, was experimentally confirmed by
Dehmelt and co-workers.
\section{Levels of Physics}
We now return to the relation (\ref{e5}) or (\ref{e20}) which expresses the
underlying non-commutative geometry of space-time.
What we would like to point out is that we are seeing here different levels of
physics. Indeed, rewriting (\ref{e5}) or (\ref{e20}) as,
$$[x,u_x] = \imath [l+l^3],$$
we can see that if $l=0$, we have classical physics, while if $0(l^3)=0,$ we have
Quantum Mechanics and finally if $0(l^3)\ne 0$ we have the above discussed
fractal picture, and from another point of view, the superstring picture.\\
Interestingly, in our case the electron Compton wavelength $l \sim 10^{-11}cm,$ so
that $0(l^3)\sim 10^{-33}$ as in string theory.\\
The expansion in terms of $l$ given above can be continued\cite{r39}, and thus
one could in principle go into deeper levels as well.
\section{Gravitation and Black Holes}
In our formulation we have not invoked the full non linear Theory of General
Relativity. General Relativity itself comes up as an approximation, in its linear
version and also through the fact that while $G$ the gravitational constant,
varies with time, over intervals small compared to the age of the universe,
it is approximately constant. (Dirac however reconciles the variation of $G$
with General Relativity by invoking the so called gravitational units of
measurement\cite{r40}, the units of our common usage
being the atomic units). The question arises, is it possible to accommodate
Black Holes within such a non General Relativistic formulation? We will now
show that Black Holes could also be understood without invoking General
Relativity at all.\\
We start by defining a Black Hole as an object at the surface of which, the
escape velocity equals the maximum possible velocity in the universe viz.,
the velocity of light. We next use the well known equation of Keplerian
orbits\cite{r41},
\begin{equation}
\frac{1}{r} = \frac{GM}{L^2}(1+e cos \theta)\label{e39}
\end{equation}
where $L$, the so called impact parameter is given by, $Rc$, where $R$ is the
point of closest approach, in our case a point on the surface of the object
and $c$ is the velocity of approach, in our case the velocity of light.\\
Choosing $\theta = 0$ and $e \approx 1$, we can deduce from (\ref{e39})
\begin{equation}
R = \frac{2GM}{c^2}\label{e40}
\end{equation}
Equation (\ref{e40}) gives the Schwarzchild radius for a Black Hole and can be
deduced from the full General Relativistic theeory.\\
We will now use (\ref{e40}) to exhibit Black Holes at three different scales,
the micro, the macro and the cosmic scales.\\
Our starting point is the observation that a Planck mass, $10^{-5}gms$ at the
Planck length, $10^{-33}cms$ satisfies (\ref{e40}) and, as such is a Schwarzchild
Black Hole. As pointed out Rosen has used non-relativistic Quantum
Theory to show that such a particle is a mini universe.\\
We next come to stellar scales. It is well known that for an electron gas in a
highly dense mass we have\cite{r42}
\begin{equation}
K\left(\frac{\bar M^{4/3}}{\bar R^4} - \frac{\bar M^{2/3}}{\bar R^2}\right)
= K' \frac{\bar M^2}{\bar R^4}\label{e41}
\end{equation}
where
\begin{equation}
\left(\frac{K}{K'}\right) = \left(\frac{27\pi}{64\alpha}\right) \left(\frac{\hbar c}
{\gamma m_P^2}\right) \approx 10^{40}\label{e42}
\end{equation}
and
$$\bar M = \frac{9\pi}{8}\frac{M}{m_P} \quad \bar R = \frac{R}{(\hbar /m_ec)},$$
$M$ is the mass, $R$ the radius of the body, $m_P$ and $m_e$ are the proton
and electron masses and $\hbar$ is the reduced Planck Constant. From
(\ref{e41}) and (\ref{e42}) it is easy to see that for $\bar M < 10^{60}$, there
are highly condensed planet sized stars.
(Infact these considerations lead to the Chandrasekhar limit in stellar
theory). We can also verify that for
$\bar M$ approaching  $10^{60}$ corresponding to a mass $\sim 10^{36}gms$, or
roughly a hundred to a thousand times the solar mass, the radius $R$ gets
smaller and smaller and would be $\sim 10^{8}cms$, so as to satisfy (\ref{e40})
and give a Black Hole in broad agreement with theory. (On 13th Septembe,
2000, NASA announced the discovery of exactly such Black Holes.)\\
Finally for the universe as a whole, using only the theory of Newtonian gravitation,
it is well known that we can deduce, as we saw in Chapter 7,
\begin{equation}
R \sim \frac{GM}{c^2}\label{e43}
\end{equation}
where this time $R \sim 10^{28}cms$ is the radius of the universe and $M \sim 10^{55}gms$
is the mass of the universe.\\
Equation (\ref{e43}) is the same as (\ref{e40}) and suggests that the universe itself is a Black Hole. It is
remarkable that if we consider the universe to be a Schwarzchild Black Hole
as suggested by (\ref{e43}), the time taken by a ray of light to traverse the
universe equals the age of the universe $\sim 10^{17}secs$ as shown elsewhere
\cite{r43}.
\section{Dimensionality and the Field and Particle Approach}
In a recent paper, Castro, Granik and El Naschie have given a rationale for
the three dimensionality of our physical space within the framework of a Cantorian
fractal space time and El Naschie's earlier work thereon\cite{r44}. An ensemble
is used and the value for the average dimension involving the golden mean is
deduced close to the value of our $3 + 1$ dimensions. We now make a few remarks
based on an approach which is in the spirit of the above considerations.\\
Our starting point is the fact that the fractal dimension of a quantum path
is two, which, it has been argued in Chapter 6 is described by the coordinates $(x, ict)$.
Infact this lead to the Dirac equation of the spin half
electron.  Given the
spin half, it was pointed out that  it is then possible to deduce the dimensionality of an ensemble
of such particles, which turns out to be three.\\
There is another way of looking at this. If we generalise from the one space
dimensional case and the complex $(x,\imath t)$ plane to three dimensions, we infact
obtain the four dimensional case and the Theory of Quarternions, which are based
on the Pauli Spin Matrices\cite{r28}. As has been noted by Sachs, had Hamilton
identified the fourth coordinate in the above generalisation with time, then
he would have anticipated Special Relativity itself. It must be observed that
the Pauli Spin Matrices which denote the Quantum Mechanical spin half form,
again, a non commutative structure.\\
Curiously enough the above consideration in the complex plane can have an
interesting connection with an unproven nearly hundred year old conjecture of
Poincare.\\
Poincare had conjectured that the fact that
closed loops could be shrunk to points on a two dimensional surface topologically
equivalent to the surface of a sphere can be generalised to three dimensions
also\cite{r45}. After all these years the conjecture has remained unproven.
We will now see why the three dimensional generalisation is not possible.\\
We firstly observe that a two dimensional surface on which closed smooth
loops can be shrunk continuously to arbitrarily small sizes is simply connected.
On such a surface we can define complex coordinates following the hydrodynamical
route exploiting the well known connection between the two. If we consider
laminar motion of an incompressible fluid we will have\cite{r46}
\begin{equation}
\vec \nabla \cdot \vec V = 0\label{e44}
\end{equation}
Equation (\ref{e44}) defines, as is well known, the stream function $\psi$ such
that
\begin{equation}
\vec V = \vec \nabla \times \psi \vec e_z\label{e45}
\end{equation}
where $\vec e_z$ is the unit vector in the $z$ direction.\\
Further, as the flow is irrotational, as well, we have
\begin{equation}
\vec \nabla \times \vec V = 0\label{e46}
\end{equation}
Equation (\ref{e46}) implies that there is a velocity potential $\phi$ such that,
\begin{equation}
\vec V = \vec \nabla \phi\label{e47}
\end{equation}
The equations (\ref{e45}) and (\ref{e47}) show that the functions $\psi$ and $\phi$
satisfy the Cauchy-Reimann equations of complex analysis\cite{r47}.\\
So it is possible to characterise the fluid elements by a complex variable
\begin{equation}
z = x + \imath y\label{e48}
\end{equation}
The question is can we generalise equation (\ref{e48}) to three dimensions? Infact
as we saw a generalisation leads not to
three but to four dimensions, with the three Pauli spin matrices $\vec \sigma$
replacing $\imath$. Further these Pauli spin matrices do not commute, and
characterise spin or vorticity. This close connection can be established by
other arguments as well\cite{r48}.\\
This is not surprising - the reason lies in equation (\ref{e45}) or equivalently
in the multiplication law of complex numbers. (Infact, there is a general
tendency to loverlook this fact and this leads to the mistaken impression that
complex numbers are just an ordered pair of numbers, which latter are
usually associated with vectors.)\\
The above considerations give an explanation for the $3 + 1$ dimensionality
of space time\cite{r49}. Moreover equations like (\ref{e45}) and (\ref{e48})
re-emphasize the hydrodynamical model discussed earlier. Incidentally as Barrow
\cite{r50} puts it, "Interestingly, the number of dimensions of space which we
experience in the large plays an important role....  It also ensures that wave phenomena behave in a coherent fashion. Were
there four dimensions of space, then simple waves would not travel at one speed
in free space, and hence we would simultaneously receive waves that were emitted
at different times. Moreover, in any world but one having three large dimensions
of space, waves would become distorted as they travelled. Such reverberation
and distortion would render any high-fidelity signalling impossible. Since so
much of the physical universe, from brain waves to quantum waves, relies upon
travelling waves we appreciate the key role played by the dimensionality of
our space in rendering its contents intelligible to us."\\
We make a final remark. We saw in Chapters 1 and 2 that while the contemporary
Field approach is based on guage interactions and spin 1 Bosons, these Bosons,
as seen in Chapter 9 are not the Quantuzed Vortices, but rather their bound
states - they can be thought of as, approxmately steamlines. On the other hand,
our approach has been based on Fermions, spin half particles, which are like
the Quantized Vortices encountered in Chapter 3.

\end{document}